\documentclass{article}

\usepackage{arxiv}

\usepackage[utf8]{inputenc} 
\usepackage[T1]{fontenc}    
\usepackage{hyperref}       
\usepackage{url}            
\usepackage{booktabs}       
\usepackage{amsfonts}       
\usepackage{nicefrac}       
\usepackage{microtype}      
\usepackage{lipsum}		
\usepackage{graphicx}
\usepackage{natbib}
\usepackage{doi}
\usepackage{amsmath}
\usepackage[ruled,vlined]{algorithm2e}
\usepackage{multirow}

\SetKw{Continue}{continue}
\SetKw{Or}{or}
\SetKw{And}{and}
\SetKw{Not}{not}

\title{Fast Matrix Multiplication via Ternary Meta Flip Graphs}

\author{\href{https://orcid.org/0000-0001-8047-0114}{\includegraphics[scale=0.06]{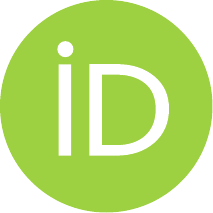}\hspace{1mm}Andrew I.~Perminov}\\
	Research Center for TAI\\
	Institute for System Programming\\
	Moscow \\
	\texttt{perminov@ispras.ru}
}

\hypersetup{
pdftitle={Fast Matrix Multiplication via Ternary Meta Flip Graphs},
pdfsubject={cs},
pdfauthor={Andrew I.~Perminov},
pdfkeywords={Fast matrix multiplication, Flip graph, Ternary integer coefficient set, Tensor rank},
}

\begin{document}
\maketitle

\begin{abstract}
Matrix multiplication optimization remains a fundamental challenge in computational mathematics. This work introduces a novel approach that discovers matrix multiplication schemes whose coefficients are restricted to the set $\{-1, 0, 1\}$ (denoted $Z_T$), minimizing naive additive complexity for efficient hardware implementation.  The core of the method is a GPU-accelerated meta flip graph algorithm that maintains ternary safety through specialized arithmetic operations and sign symmetry breaking. Key results include new best ranks for the formats $4 \times 5 \times 12$, $5 \times 6 \times 10$, and $6 \times 7 \times 9$, the independent discovery of 32 schemes in $Z_T$ that match known optimal ranks (including 8 previously known only with rational coefficients), and 30 rank improvements in the binary field. The analysis of 164 known schemes shows that 92 admit a ternary-coefficient implementation, while 72 could not be found under this constraint, defining the current boundaries of the approach. All software, results, and discovered schemes are provided as open-source.
\end{abstract}


\section{Introduction}
Matrix multiplication is a core operation in computing, used in areas like scientific simulation, machine learning, and graphics. The search for faster methods began with Strassen’s 1969 breakthrough, which showed that two $2 \times 2$ matrices could be multiplied using only 7 multiplications instead of 8~\citep{strassen1969gaussian}. This result opened a new research field focused on reducing the number of multiplications needed for matrix multiplication.

Recent years have seen automated approaches for discovering efficient algorithms. SAT solving encodes the problem as Boolean satisfiability, finding optimal schemes for small matrices but facing scalability limits~\citep{heule2019local}. The flip graph approach~\citep{kauers2023flip} treats algorithms as graph vertices connected by flip operations, with subsequent improvements adding adaptive search~\citep{arai2024adaptive} and meta operations across dimensions~\citep{kauers2025exploring}.

However, a key limitation remains: most automated search tools work only in the binary field ($Z_2$), where coefficients are only 0 or 1. While fast to search, these schemes often cannot be converted to practical integer ($Z$) or rational ($Q$) coefficients without introducing large numbers that increase computation cost. Other approaches, including human designs, SAT solving, and AI methods like DeepMind's AlphaTensor~\citep{fawzi2022discovering} and AlphaEvolve~\citep{novikov2025alphaevolve}, often find algorithms that use coefficients like 2, 4, 9, or fractions and complex numbers, which are inefficient in real hardware.

This work focuses on matrix multiplication schemes where all coefficients are restricted to the set $Z_T = \{-1, 0, 1\}$, referred to as ternary coefficients. This approach should not be confused with working over the finite field $\mathbb{F}_3$. Instead, it constrains coefficients for hardware efficiency while maintaining schemes over $\mathbb{Q}$ or $\mathbb{R}$. This avoids costly multiplications by large numbers while keeping algorithms efficient. A new GPU-accelerated search algorithm was built to explore only ternary schemes, using custom arithmetic and sign symmetry breaking rules. The system also includes new operations -- like project, extend, merge, product, and size swap -- to efficiently explore the space of possible algorithms.

The main contributions of this work are:

\begin{itemize}
    \item GPU-based tool for finding matrix multiplication schemes using only $\{-1, 0, 1\}$ coefficients.

    \item New record-breaking ranks for formats $4 \times 5 \times 12$, $5 \times 6 \times 10$, and $6 \times 7 \times 9$ formats, and reduced naive addition complexity for 40 ternary schemes.
    
    \item Rediscovery of 32 schemes using ternary coefficients ($Z_T = \{-1, 0, 1\}$) that match the best-known ranks, including 8 that were previously only known with rational coefficients ($Q$), and 30 rank improvements in the binary field ($Z_2$).

    \item Open-source release of all software tools and results on GitHub.
\end{itemize}

The paper is organized as follows: section \ref{sec:related_works} reviews relevant background and related work. Section \ref{sec:methodology} details the methodology and algorithmic innovations. Section~\ref{seq:experimental_results} presents experimental results, Section~\ref{seq:discussion} discusses the findings and Section~\ref{seq:conclusion} concludes with directions for future research.

This paper uses the following notation for matrix multiplication schemes:

\begin{itemize}
    \item A scheme denoted as $(m, n, p: r)$ describes an algorithm for multiplying an $m \times n$ matrix by an $n \times p$ matrix using $r$ non-scalar multiplications.

    \item The value $r$ is referred to as the rank of the scheme.

    \item The \textit{naive addition complexity} counts the total number of addition and subtraction operations needed to compute the linear combinations in the scheme, without any optimization for common subexpressions.

   \item The coefficient sets and rings are denoted as:
    \begin{itemize}
        \item $Z_T = \{-1, 0, 1\}$ -- the ternary coefficient set;
        \item $Z_2 = {0, 1}$ -- the binary field;
        \item $Z$ -- the ring of integers (coefficients beyond $\{-1, 0, 1\}$);
        \item $Q$ -- the field of rational numbers.
    \end{itemize}
\end{itemize}

This notation distinguishes schemes using only simple ternary coefficients ($Z_T$) from those requiring arbitrary integers ($Z$) or rational numbers ($Q$).

\section{Background and Related Work}
\label{sec:related_works}

\subsection{Matrix Multiplication Complexity}
Since Strassen's 1969 breakthrough, researchers have known that standard $O(n^3)$ matrix multiplication is not optimal. Strassen showed that two $2 \times 2$ matrices could be multiplied using only 7 multiplications instead of 8. This started the search for better algorithms.

While recent theoretical work has improved the asymptotic complexity to $O(n^{2.371339})$~\citep{alman2025more}, these methods have large constant factors that make them impractical for real use. For small matrices, researchers now focus on finding the exact minimum number of multiplications needed -- called the rank -- for specific matrix sizes like $3 \times 3$ or $4 \times 4$. These small algorithms are actually useful in real applications like deep learning and scientific computing.

\subsection{Algorithm Discovery Approaches}
Various automated methods have been developed to tackle the problem of discovering matrix multiplication algorithms.

\subsubsection{SAT Solving}
SAT solvers have been applied by encoding the problem as Boolean satisfiability instances. For small formats like $(2, 2, 2: 7)$, SAT solvers can find solutions in seconds. However, for $(3, 3, 3: 23)$ format, exhaustive SAT solving becomes infeasible, running for days without finding solutions.

Recent work has improved SAT-based approaches through symmetry breaking techniques, which help reduce the search space by eliminating redundant solutions~\citep{yang2024ruling}. The practical solution also combines SAT with local search heuristics. By starting from randomly initialized coefficients, researchers have discovered more than 17,000 non-equivalent $(3, 3, 3: 23)$ schemes~\citep{heule2019local, heule2021new}. This hybrid approach shows that while pure SAT solving faces scalability limits, guided search with SAT components remains valuable.

\subsubsection{Reinforcement Learning}
DeepMind's AlphaTensor~\citep{fawzi2022discovering} demonstrated that deep reinforcement learning can discover novel matrix multiplication algorithms. It frames algorithm discovery as a single-player game where the objective is to factorize the matrix multiplication tensor. The subsequent AlphaEvolve work~\citep{novikov2025alphaevolve} uses language models to further advance automated discovery.

However, these AI-discovered algorithms often employ coefficients that are suboptimal for practical implementation, such as large integers, fractions or complex numbers, which introduce significant computational overhead in hardware implementations.

\subsubsection{Constraint Programming}
Constraint programming offers a declarative approach to finding matrix multiplication schemes. By modeling the Brent equations as a constraint satisfaction problem, solvers can search for valid coefficient assignments. This method has been used to discover schemes from scratch for small formats~\citep{deza2023fast}, demonstrating its effectiveness for exhaustive search in constrained problem spaces.

\subsubsection{Flip Graph Methods}
The flip graph approach, introduced by Moosbauer et al.~\citep{kauers2023flip}, reformulates the search as a graph exploration problem. In this model, each vertex represents a valid matrix multiplication scheme, and edges correspond to flips -- local transformations that modify the scheme while preserving its correctness. This combinatorial perspective enables systematic exploration of the algorithm space.

Subsequent work introduced an adaptive flip graph algorithm with a plus operator~\citep{arai2024adaptive}, which enhanced search efficiency. The most recent advancement, the meta flip graph methodology~\citep{kauers2025exploring}, extends this approach by enabling transitions between different matrix dimensions, significantly expanding the searchable space. This work builds directly upon the meta flip graph paradigm.

\subsection{Coefficient Rings}
The choice of number system (called a coefficient ring) greatly affects both the discovery and practicality of matrix multiplication algorithms. Most useful algorithms need integer or rational coefficients, but the fastest search methods work in the binary field ($Z_2$).

This creates a problem: algorithms found with binary field search often can't be converted to practical integer coefficients. The FMM Catalogue~\citep{sedoglavic2023yet} stores schemes with the best-known ranks, but sometimes it's better to store algorithms in different coefficient rings that are more suitable for hardware implementation, even if they have slightly higher ranks.

\subsection{Additive Complexity}
Besides multiplication count, the number of additions and subtractions (additive complexity) matters greatly in practice. Algorithms with minimal multiplications but many additions can be slower in real hardware~\citep{brent1970algorithms}. This work focuses on finding algorithms that minimize both multiplications and naive additions by using only simple coefficients: $-1$, $0$, and $1$. The approach maximizes the number of zero coefficients in the scheme tensors, directly reducing the naive addition count. This provides a foundation for further optimization through common subexpression elimination~\citep{maartensson2025number}, which can significantly reduce the actual number of operations in implementation. A detailed treatment of advanced addition minimization techniques will be presented in a separate study to maintain focus on the core rank optimization problem.

\section{Methodology}
\label{sec:methodology}

\subsection{Matrix Multiplication Schemes}
\label{subsec:brent_equations}

A matrix multiplication scheme for multiplying matrices $A \in \mathbb{F}^{m \times n}$ and $B \in \mathbb{F}^{n \times p}$ over an arbitrary field $\mathbb{F}$ with rank $r$ consists of three sets of coefficients that define the computation - $u^{(l)}_{ij}$, $v^{(l)}_{ij}$ and $w^{(l)}_{ij}$. The intermediate products (multiplications) are computed as:

\begin{align*}
m_1 = (u^{(1)}_{11} a_{11} + \cdots + u^{(1)}_{mn} a_{mn}) \cdot & (v^{(1)}_{11} b_{11} + \cdots + v^{(1)}_{np} b_{np})\\
\vdots \\
m_r = (u^{(r)}_{11} a_{11} + \cdots + u^{(r)}_{mn} a_{mn}) \cdot & (v^{(r)}_{11} b_{11} + \cdots + v^{(r)}_{np} b_{np}),
\end{align*}

and the elements of the result matrix $C = AB$ are calculated as:

\begin{align*}
c_{ij} = w^{(1)}_{ij}m_1 + \cdots + w^{(r)}_{ij}m_r.
\end{align*}

In this formulation, the coefficients are organized into three tensors: $U \in \mathbb{F}^{r \times m \times n}$, $V \in \mathbb{F}^{r \times n \times p}$, and $W \in \mathbb{F}^{r \times m \times p}$. The scheme computes $r$ intermediate products, each being a linear combination of entries from $A$ multiplied by a linear combination of entries from $B$, followed by recombination using coefficients from $W$ to produce the final result.

Comparing the coefficients of all terms $a_{i_1 i_2}$, $b_{j_1 j_2}$, $c_{k_1 k_2}$ in the equations $c_{ij} = \sum_k{a_{ik}b_{kj}}$ leads to the polynomial equations also known as Brent equations~\citep{brent1970algorithms}:

\begin{equation}
\label{eq:brent_eq}
    \sum\limits_{l=1}^r {u^{(l)}_{i_1i_2} v^{(l)}_{j_1j_2}w^{(l)}_{k_1k_2}} = \delta_{i_2j_1} \delta_{i_1k_1} \delta_{j_2k_2}
\end{equation}
for $i_1, k_1 \in \{1, \cdots m\}$, $i_2, j_1 \in \{1, \cdots, n\}$ and $j_2,k_2 \in \{1, \cdots, p\}$. The $\delta_{ij}$ on the right are Kronecker-deltas,
i.e., $\delta_{ij} = 1$ if $i = j$ and $\delta_{ij} = 0$ otherwise.

For mathematical symmetry and computational convenience, the formulation uses $C^T$ rather than $C$ directly, which yields more symmetric Brent equations during the search process. In this representation, the tensors have dimensions $U \in \mathbb{F}^{r \times m \times n}$, $V \in \mathbb{F}^{r \times n \times p}$ and $W \in \mathbb{F}^{r \times p \times m}$, and the elements are computed as:

\begin{align*}
c_{ji} = w^{(1)}_{ij}m_1 + \cdots + w^{(r)}_{ij}m_r.
\end{align*}

This representation aligns with standard practice in the matrix multiplication algorithm literature and simplifies the constraint satisfaction problem~\citep{heule2019local, deza2023fast, yang2024ruling}.

\subsubsection{Example: a 2×2 by 2×3 Matrix Multiplication Scheme}
The matrices $U$, $V$ and $W$ below represent a $(2, 2, 3: 11)$ scheme. Each row corresponds to the coefficients for one of the 11 intermediate multiplications $m_1, \dots, m_{11}$.

\[
U = \begin{bmatrix}
    1 & 0 & 0 & 1 \\
    1 & 0 & 0 & 0 \\
    0 & 0 & 0 & 1 \\
    -1 & 1 & 0 & 0 \\
    0 & 1 & 0 & 1 \\
    1 & 0 & 1 & 0 \\
    0 & 0 & 1 & -1 \\
    1 & 0 & 0 & 0 \\
    0 & 1 & 0 & 0 \\
    0 & 0 & 1 & 0 \\
    0 & 0 & 0 & 1
\end{bmatrix}
\quad
V = \begin{bmatrix}
    1 & 0 & 0 & 0 & 1 & 0 \\
    0 & 1 & 0 & 0 & 1 & 0 \\
    1 & 0 & 0 & 1 & 0 & 0 \\
    0 & 0 & 0 & 0 & 1 & 0 \\
    0 & 0 & 0 & 1 & -1 & 0 \\
    -1 & 1 & 0 & 0 & 0 & 0 \\
    1 & 0 & 0 & 0 & 0 & 0 \\
    0 & 0 & 1 & 0 & 0 & 0 \\
    0 & 0 & 0 & 0 & 0 & 1 \\
    0 & 0 & 1 & 0 & 0 & 0 \\
    0 & 0 & 0 & 0 & 0 & 1
\end{bmatrix}
\quad
W = \begin{bmatrix}
    1 & 0 & 0 & 1 & 0 & 0 \\
    0 & 0 & 1 & -1 & 0 & 0 \\
    -1 & 1 & 0 & 0 & 0 & 0 \\
    1 & 0 & 1 & 0 & 0 & 0 \\
    1 & 0 & 0 & 0 & 0 & 0 \\
    0 & 0 & 0 & 1 & 0 & 0 \\
    0 & 1 & 0 & 1 & 0 & 0 \\
    0 & 0 & 0 & 0 & 1 & 0 \\
    0 & 0 & 0 & 0 & 1 & 0 \\
    0 & 0 & 0 & 0 & 0 & 1 \\
    0 & 0 & 0 & 0 & 0 & 1

\end{bmatrix}
\]

The matrices use flattened indices:
\begin{itemize}
    \item for $U$ columns represent $(a_{11}, a_{12}, a_{21}, a_{22})$;
    \item for $V$ columns represent $(b_{11}, b_{12}, b_{13}, b_{21}, b_{22}, b_{23})$;
    \item for $W$ columns represent $(c_{11}, c_{21}, c_{12}, c_{22}, c_{13}, c_{23})$.
\end{itemize}

For instance, the first row of $U$ and $V$ gives
\[
    m_1=(a_{11} + a_{22}) \cdot (b_{11} + b_{22}),
\]

and the first column of $W$ shows how $c_{11}$ is computed:
\[
    c_{11} = m_1 - m_3 + m_4 + m_5.
\]

This demonstrates how the scheme combines matrix elements and reconstructs the result through linear combinations of the intermediate products.

\subsection{The Ternary Coefficients Approach}
This work explores matrix multiplication schemes restricted to the ternary coefficient set $Z_T = \{-1, 0, 1\}$, which offer practical advantages for hardware implementation. While schemes with larger coefficients (2, 4, 9, etc.) or fractions may achieve theoretically optimal ranks, they introduce substantial additive complexity that reduces their practical utility. The ternary constraint enables efficient hardware representation using minimal resources, eliminating the need for expensive multiplication or division operations during coefficient application.

In this system, all coefficients in the tensors $U$, $V$, and $W$ are constrained to $Z_T$, ensuring the entire computation requires only additions and subtractions without scalar multiplications by coefficients other than $\pm 1$. This approach maintains the mathematical structure described in Section~\ref{subsec:brent_equations} while imposing the practical constraint that all coefficients must be in $\{-1, 0, 1\}$.

\subsection{Algorithmic Operators}
The search methodology employs several operators for transforming and combining matrix multiplication schemes while maintaining mathematical correctness. These operators form the foundation of the flip graph approach and enable local exploration of the algorithm space.

\subsubsection{Core Operators}

\begin{itemize}
    \item \textbf{Flip}: for two rank-one tensors satisfying $u^{(i)} = u^{(j)}$ the transformation is defined as:
    \begin{align*}
        u^{(i)} \otimes v^{(i)} \otimes w^{(i)} \quad+&\quad u^{(j)} \otimes v^{(j)} \otimes w^{(j)}
        \rightarrow \\
        u^{(i)} \otimes (v^{(i)} + v^{(j)}) \otimes w^{(i)} \quad+&\quad u^{(j)} \otimes v^{(j)} \otimes (w^{(j)} - w^{(i)})
    \end{align*}

    This operation preserves the scheme's rank while modifying its structure.

    \item \textbf{Plus}: for two rank-one tensors satisfying $u^{(i)} \neq u^{(j)}$, $v^{(i)} \neq v^{(j)}$, and $w^{(i)} \neq w^{(j)}$, the transformation expands the scheme as:
    \begin{align*}
        u^{(i)} \otimes v^{(i)} \otimes w^{(i)} \quad+&\quad u^{(j)} \otimes v^{(j)} \otimes w^{(j)} \quad \rightarrow\\
u^{(i)} \otimes (v^{(i)} + v^{(j)}) \otimes w^{(i)} \quad+&\quad u^{(i)} \otimes v^{(j)} \otimes (w^{(j)} - w^{(i)}) \quad+\quad (u^{(j)} - u^{(i)}) \otimes v^{(j)} \otimes w^{(j)}
    \end{align*}

    This operation increases the scheme's rank while preserving correctness.

    \item \textbf{Split}: for two rank-one tensors satisfying $u^{(i)} \neq u^{(j)}$, the transformation expands the scheme as:
    \begin{align*}
        u^{(i)} \otimes v^{(i)} \otimes w^{(i)} \quad+&\quad u^{(j)} \otimes v^{(j)} \otimes w^{(j)} \quad \rightarrow\\
        u^{(j)} \otimes v^{(i)} \otimes w^{(i)} \quad+&\quad u^{(j)} \otimes v^{(j)} \otimes w^{(j)} \quad+\quad (u^{(i)} - u^{(j)}) \otimes v^{(i)} \otimes w^{(i)}
    \end{align*}

    This also increases the rank while maintaining correctness.

    \item \textbf{Reduction}: for two rank-one tensors satisfying $u^{(i)} = u^{(j)}$ and $v^{(i)} = v^{(j)}$, the transformation combines them as:
    \begin{align*}
        u^{(i)} \otimes v^{(i)} \otimes w^{(i)} \quad+&\quad u^{(j)} \otimes v^{(j)} \otimes w^{(j)} \quad \rightarrow \quad  u^{(i)} \otimes v^{(i)} \otimes (w^{(i)} + w^{(j)})
    \end{align*}

    This operation decreases the scheme's rank by eliminating redundant components.
\end{itemize}

All these operators can be applied to any permutation of $u$, $v$, and $w$ tensors, providing comprehensive coverage of possible local transformations. The functionality of the \texttt{plus} and \texttt{split} operators is combined into a single operator named \texttt{expand}. During execution, it selects uniformly at random whether to apply a \texttt{plus} or a \texttt{split} transformation.

\subsubsection{Meta Operators}
\begin{itemize}
    \item \textbf{Project}: transforms a scheme from format $(m, n, p)$ to $(m, n, p - 1)$ by excluding one dimension and remove zero terms.

    \item \textbf{Extend}: converts a scheme from format $(m, n, p: r)$ to $(m, n, p + 1: r + mn)$ by adding a naive $(m, n, 1)$ scheme.

    \item \textbf{Merge}: combines two schemes $(m, n, p_1: r_1)$ and $(m, n, p_2: r_2)$ into $(m, n, p_1 + p_2: r_1 + r_2)$.

    \item \textbf{Double}: merges a scheme with itself to double one dimension.

    \item \textbf{Product}: computes the tensor product of two schemes $(m_1, n_1, p_1: r_1)$ and $(m_2, n_2, p_2: r_2)$ to yield $(m_1 \cdot m_2, n_1\cdot n_2, p_1 \cdot p_2: r_1 \cdot r_2)$ scheme.

    \item \textbf{Swap sizes}: transforms $(m, n, p: r)$ to $(m, p, n: r)$ through matrix transposition operations.
\end{itemize}

These meta operators enable exploration across different matrix dimensions, allowing the algorithm to transfer knowledge between formats and construct complex schemes from simpler building blocks.

\subsubsection{Advanced Composition}
\begin{itemize}
    \item \textbf{Block matrix multiplication}: constructs larger schemes using 2×2 block matrix decomposition with small existing schemes, particularly employing the Strassen 2×2×2 scheme as a building block.
\end{itemize}

\subsection{FlipGraphGPU Architecture}
\subsubsection{Overall Algorithm Structure}

The FlipGraphGPU algorithm operates through an iterative process that maintains a population of schemes and applies transformations to discover improved variants:

\begin{enumerate}
    \item \textbf{Initialization}: create $N$ initial schemes by either generating naive implementations or loading existing schemes from disk. If fewer than $N$ schemes are available, existing schemes are duplicated to fill the population.

    \item \textbf{Rank assessment}: calculate current best ranks for every unique matrix format $(m \times n \times p)$ across the population, establishing baseline performance metrics.

    \item \textbf{Random walk}: execute the RandomWalk kernel for `maxIterations` iterations. During each iteration:
    \begin{itemize}
        \item Apply the \texttt{flip} operation with reduction edge checking;
        \item Evaluate rank improvement, storing new best schemes when discovered;
        \item Probabilistically apply \texttt{expand} operator.
    \end{itemize}

    \item \textbf{Resize}: execute the Resize kernel to explore dimensional transformations:

    \begin{itemize}
        \item Randomly swap scheme dimensions $(m, n, p)$;
        \item Attempt to \texttt{merge} with randomly selected best schemes;
        \item Apply \texttt{project}, \texttt{product}, \texttt{double} and \texttt{extend} operators with defined probabilities.
    \end{itemize}

    \item \textbf{Population synchronization}: re-evaluate unique scheme dimensions $(m \times n \times p)$ across the entire population and update best scheme mappings. This step is crucial since the resize phase alters scheme dimensions, making previous best scheme references potentially obsolete.
\end{enumerate}

This cyclic process of local optimization (random walk) followed by dimensional exploration (resize) enables comprehensive exploration of both the algorithm space for fixed dimensions and the space of possible matrix formats. The persistent storage of improvements ensures that discoveries are retained across algorithm restarts, while the population-based approach maintains diversity in the search process.

The complete implementation is publicly available at \url{https://github.com/dronperminov/FlipGraphGPU}.

\subsubsection{RandomWalk Kernel}
The RandomWalk Kernel explores the local neighborhood of each scheme while keeping matrix dimensions fixed. Each scheme is processed independently with its own random number generator for consistent results, as shown in Algorithm~\ref{alg:random_walk}.

\begin{algorithm}[ht!]
    \label{alg:random_walk}
    \KwIn{scheme, best\_scheme, best\_rank, max\_iterations, $p_{reduce}$, $p_{expand}$}
    \KwOut{updated scheme, best\_scheme, best\_rank}
    \For{$iteration \gets 1$ \KwTo $max\_iterations$}{
        \If {\Not $scheme.try\_flip()$}{
            scheme.expand()\;
            \Continue\;
        }
        \BlankLine
        \If {$scheme.rank < best\_rank$ \Or $scheme.rank = best\_rank$ \And $random() < 0.01$}{
            $best\_rank \gets scheme.rank$\;
            $best\_scheme \gets scheme$\;
        }
        \BlankLine
        \If {$random() < p_{reduce}$}{
            scheme.reduce()\;
        }
        \BlankLine
        \If{$random() < p_{expand}$ \And $scheme.rank \leq best\_rank + 2$}{
            scheme.expand()\;
        }
    }
    \Return {updated scheme, best\_scheme, best\_rank}
    \caption{RandomWalk kernel pseudocode}
\end{algorithm}

Key features of the kernel include:

\begin{itemize}
    \item \textbf{Adaptive acceptance}: better-ranked schemes are always saved, while equal-ranked schemes are occasionally accepted (1\% probability) to escape local optima.

    \item \textbf{Controlled exploration}: the \texttt{expand} operator is restricted to schemes close to the current best rank to prevent excessive rank growth.

    \item \textbf{Probabilistic operations}: \texttt{reduce} and \texttt{expand} operations are applied with predefined probabilities to balance exploration and exploitation.
\end{itemize}

\subsubsection{Resize Kernel}
The resize kernel explores different matrix dimensions to discover algorithms across various matrix sizes. The approach is detailed in Algorithm~\ref{alg:resize}.

\begin{algorithm}[ht!]
    \label{alg:resize}
    \KwIn{scheme, best\_schemes, $p_{resize}$}
    \KwOut{updated scheme}
    \If{random() < 0.5}
        {scheme.swap\_sizes()}
    \BlankLine
    random\_scheme $\gets$ random\_select(best\_schemes)\;
    merged $\gets$ scheme.try\_merge(random\_scheme)\;
    \BlankLine
    \If {\Not merged \And random() < $p_{resize}$}{
        p $\gets$ random()\;
        \BlankLine
        \If {p < 0.05}{
            scheme.project()\;
        }
        \ElseIf {p < 0.55}{
            random\_scheme $\gets$ random\_select(best\_schemes)\;
            scheme.product(random\_scheme)\;
        }
        \ElseIf {p < 0.85}{
            scheme.double()\;
        }
        \Else {
            scheme.extend()\;
        }
    }
    \Return {updated scheme}
    \caption{Resize kernel pseudocode}
\end{algorithm}

Key strategies in the resize kernel include:

\begin{itemize}
    \item \textbf{Dimension flexibility}: random size swapping explores alternative dimension orderings.

    \item \textbf{Collaborative merging}: attempting to merge with best schemes enables knowledge transfer.

    \item \textbf{Balanced operator selection}: the probability distribution (5\% project, 50\% product with best schemes, 30\% double, 15\% extend) ensures comprehensive exploration.
\end{itemize}

This dual-kernel approach enables the algorithm to efficiently balance intensive local search within fixed dimensions with global exploration across the space of possible matrix formats.

\subsection{Ternary Safety Mechanisms}
Implementing operations that preserve the ternary coefficient constraint presents greater computational challenges than working in the binary field. In $Z_2$, matrix schemes can be efficiently represented using \texttt{uint64\_t} bitmasks with arithmetic handled through fast XOR operations. However, the ternary set $Z_T = \{-1, 0, 1\}$ requires representing three distinct states per coefficient instead of two, making simple \texttt{uint64\_t} representations impossible.

A straightforward approach would store ternary coefficients in arrays of small integers. However, this method incurs substantial performance penalties due to increased memory bandwidth requirements and reduced computational density. To maintain the high performance characteristics of binary field implementations while supporting ternary constraints, a specialized packed representation was developed that efficiently encodes the three coefficient states.

\subsubsection{One-Bit Number Representation}
To maintain ternary constraints while achieving high performance, vectors of ternary values are represented using a compact format with two \texttt{uint64\_t} variables:

\begin{itemize}
\item \texttt{n}: number of elements in the vector;
\item \texttt{digits}: bitmask indicating positions with non-zero values;
\item \texttt{signs}: bitmask indicating the sign (1 for negative, 0 for positive) of non-zero values;
\item \texttt{valid}: boolean flag indicating whether all values remain in $Z_T$.
\end{itemize}

This representation efficiently packs multiple ternary coefficients into bit-level operations while preserving the ability to perform arithmetic with overflow detection across the entire vector.

\subsubsection{Ternary Arithmetic Operations}
The arithmetic operations are designed to maintain ternary safety:

\textbf{Addition}:
\begin{align*}
digits_{a+b} =&\; digits_a \oplus digits_b \\
signs_{a+b} =&\; (signs_a \land digits_a \lor signs_b \land digits_b) \land digits_{a+b} \\
valid_{a+b} =&\; digits_a \land digits_b \land \overline{(signs_a \oplus signs_b)} = 0
\end{align*}

\textbf{Subtraction}:
\begin{align*}
digits_{a-b} =&\; digits_a \oplus digits_b \\
signs_{a-b} =&\; (signs_a \land digits_a \lor \overline{signs_b} \land digits_b) \land digits_{a-b} \\
valid_{a-b} =&\; digits_a \land digits_b \land (signs_a \oplus signs_b) = 0
\end{align*}

\textbf{Comparison}:
\begin{align*}
    a \neq b: &\; (digits_a \neq digits_b) \lor (signs_a \neq signs_b) \\
    a = b: &\; (digits_a = digits_b) \land (signs_a = signs_b) \\
    a = -b: &\; (digits_a = digits_b) \land (signs_a = (\overline{signs_b} \land digits_b))
\end{align*}

This bit-level arithmetic provides performance comparable to $Z_2$ operations while fully supporting the ternary coefficient system and detecting when operations would violate the ternary constraint.

\subsection{Sign symmetry breaking}
The ternary coefficient set introduces significant complexity in flip operations due to the presence of both positive and negative coefficients. A critical challenge arises from the need to consider both exact equality and inverse equality when identifying valid flip candidates. For example, pairs like $u^{(i)} \otimes v^{(i)} \otimes w^{(i)}$ and $u^{(j)} \otimes v^{(j)} \otimes w^{(j)}$ where $u^{(i)} = -u^{(j)}$ are mathematically valid for flip operations but require additional handling.

To address this challenge while maintaining computational efficiency, a sign symmetry breaking convention is enforced: the first non-zero coefficient in each column of the $U$ and $V$ matrices must be positive. This normalization is mathematically justified by the equivalence transformation $\alpha u \times \beta v \times \gamma w$ where $\alpha\beta\gamma = 1$, which allows coefficient rescaling without affecting the scheme's correctness.

\textbf{Example:} the following matrices demonstrate sign symmetry breaking in practice. The original scheme $(U, V, W)$ contains negative first coefficients in several columns of $U$ and $V$.

\[
    U = \begin{bmatrix}
        1 & 0 & 0 & 1 \\
        0 & -1 & 0 & 1 \\
        -1 & 0 & 1 & 0 \\
        1 & 1 & 0 & 0 \\
        1 & 0 & 0 & 0 \\
        0 & 0 & 0 & 1 \\
        0 & 0 & 1 & 1
    \end{bmatrix}
    \quad
    V = \begin{bmatrix}
        1 & 0 & 0 & 1 \\
        0 & 0 & -1 & -1 \\
        -1 & -1 & 0 & 0 \\
        0 & 0 & 0 & -1 \\
        0 & 1 & 0 & -1 \\
        -1 & 0 & 1 & 0 \\
        1 & 0 & 0 & 0
    \end{bmatrix}
    \quad
    W = \begin{bmatrix}
        1 & 0 & 0 & 1 \\
        1 & 0 & 0 & 0 \\
        0 & 0 & 0 & -1 \\
        1 & 0 & -1 & 0 \\
        0 & 0 & 1 & 1 \\
        1 & 1 & 0 & 0 \\
        0 & 1 & 0 & -1
    \end{bmatrix}
\]

After normalization, the equivalent scheme $(U', V', W')$ ensures all first non-zero coefficients in $U'$ and $V'$ are positive:

\[
    U' = \begin{bmatrix}
        1 & 0 & 0 & 1 \\
        0 & 1 & 0 & -1 \\
        1 & 0 & -1 & 0 \\
        1 & 1 & 0 & 0 \\
        1 & 0 & 0 & 0 \\
        0 & 0 & 0 & 1 \\
        0 & 0 & 1 & 1
    \end{bmatrix}
    \quad
    V' = \begin{bmatrix}
        1 & 0 & 0 & 1 \\
        0 & 0 & 1 & 1 \\
        1 & 1 & 0 & 0 \\
        0 & 0 & 0 & 1 \\
        0 & 1 & 0 & -1 \\
        1 & 0 & -1 & 0 \\
        1 & 0 & 0 & 0
    \end{bmatrix}
    \quad
    W' = \begin{bmatrix}
        1 & 0 & 0 & 1 \\
        1 & 0 & 0 & 0 \\
        0 & 0 & 0 & -1 \\
        -1 & 0 & 1 & 0 \\
        0 & 0 & 1 & 1 \\
        -1 & -1 & 0 & 0 \\
        0 & 1 & 0 & -1
    \end{bmatrix}
\]

\begin{itemize}
    \item In $U$ row 2 begins with $0, -1$ → in $U'$ row 2 becomes $0, 1$ with sign compensation in other matrices.
    \item In $U$ row 3 begins with $-1, 0$ → in $U'$ row 3 becomes $1, 0$.
    \item In $V$ multiple rows have negative first non zero coefficients that are normalized in $V'$.
\end{itemize}

The $W$ matrix absorbs the sign changes to maintain mathematical correctness, with several coefficients flipped between $W$ and $W'$. This transformation demonstrates how equivalent schemes can be normalized while preserving the algorithm's validity, enabling more efficient flip identification during search.

For the $W$ matrix, inverse equality checking is preserved since sign variations in $W$ don't break the symmetry breaking convention and may lead to valid ternary flips. This balanced approach -- strict sign normalization for $U$ and $V$ with flexibility for $W$ -- provides the necessary constraints for efficient search while maintaining expressiveness in the algorithm space.

The implementation enforces this convention during scheme initialization and after each transformation, ensuring that all schemes remain in the normalized form throughout the search process.

\subsection{Isotropy Invariants and Novelty Verification}
To characterize discovered schemes and verify they represent new tensor decompositions rather than points in known orbits under the de Groote isotropy group, three types of invariants are computed. For each scheme $(U,V,W)$ of rank $r$, the following invariants are calculated:

\begin{itemize}
    \item \textbf{Type invariant}: following the FMM Catalogue~\citep{sedoglavic2023yet}, the polynomial
    \[\sum_{i=1}^{r} X^{\text{rank} U^{(i)}} Y^{\text{rank} V^{(i)}} Z^{\text{rank}W^{(i)}}\]
    is computed, where $\text{rank}(M)$ denotes the classical matrix rank of $M$.

    \item \textbf{Symmetric polynomial invariant}: following the "3x3 Matrix Multiplication Solution Repository"~\citep{heule2019local}, the symmetrized polynomial
    \[f(x, y, z) = \sum_{\pi \in S_3} \pi \left( \sum_{i=1}^{r} x^{\text{rank}U^{(i)}} y^{\text{rank}V^{(i)}} z^{\text{rank}W^{(i)}} \right)\]
    is computed, where the permutations $\pi \in S_3$ act on the variables $x, y, z$.

    \item \textbf{Rank sum polynomial}: additionally, the polynomial
    \[g(\omega) = \omega^{\sum_{i=1}^{r} \text{rank} U^{(i)}} + \omega^{\sum_{i=1}^{r} \text{rank}V^{(i)}} + \omega^{\sum_{i=1}^{r} \text{rank}W^{(i)}}\]
    is computed.
\end{itemize}

These invariants are computed for all newly discovered schemes and compared against those of known decompositions. The complete invariant data, which is too extensive to include in the main text, is provided within the accompanying scheme files in the open-source repository. This analysis confirms that the ternary-constrained schemes discovered via the flip graph approach represent genuinely new points in the space of matrix multiplication algorithms.

\subsection{Initial Scheme Sources}
The search process was initialized using matrix multiplication schemes from established sources:

\begin{itemize}
\item FMM Catalogue~\citep{sedoglavic2023yet};
\item AlphaTensor~\citep{fawzi2022discovering};
\item AlphaEvolve~\citep{novikov2025alphaevolve};
\item Flip Graph~\citep{kauers2023flip};
\item Adaptive Flip Graph~\citep{arai2024adaptive};
\item Symmetric Flip Graph~\citep{moosbauer2025flip};
\item Meta Flip Graph~\citep{kauers2025exploring}.
\end{itemize}

From these sources, only schemes already employing ternary coefficients (${-1, 0, 1}$) were selected for initialization. This approach ensured the search process began within the feasible space of ternary-constrained implementations while building upon existing research.

\subsection{Search Strategy Variants}
The algorithm supports three main operational modes:

\begin{itemize}
    \item Focused search: operates without resize operations, using the full \texttt{maxIterations} for intensive local exploration of specific formats.

    \item Exploratory search: employs both \texttt{RandomWalk} and \texttt{Resize} kernels to discover improvements across multiple formats simultaneously.

    \item \textbf{Naive complexity minimization}: performs random flips without reduction edges to find schemes with minimal naive additive complexity, focusing on practical performance rather than just rank reduction.
\end{itemize}

This methodology enables efficient discovery of matrix multiplication schemes with ternary coefficients while maintaining the performance required for comprehensive exploration of the algorithm space.

\subsection{Supporting Tools}
The main FlipGraphGPU algorithm works together with additional tools that improve its functionality. These tools help convert, diversify, and validate matrix multiplication schemes, creating a complete research system.

\subsubsection{Ternary Lifting Script}
This tool converts binary field ($Z_2$) schemes to implementations with ternary coefficients by solving the sign assignment problem. Using OR-Tools~\citep{perron2025google} constraint programming, the tool assigns ±1 signs to non-zero coefficients while preserving Brent equation~\ref{eq:brent_eq} validity. This approach is related to the constraint programming methods used to discover matrix multiplication schemes from scratch~\citep{deza2023fast}. The implementation supports solving for multiple valid sign assignments.

\subsubsection{Alternative Scheme finder}
Discovers novel matrix multiplication schemes through SAT-based diversification. The algorithm starts from existing schemes and performs probabilistic preservation of $U$, $V$, and $W$ coefficients with configurable probabilities, followed by Brent equation solving via CryptoMiniSat 5~\citep{soos2016cryptominisat}. Incorporates random flip iterations for local exploration before SAT solving. Employs specialized XOR encoding of Brent equations with symmetry breaking via lexicographical ordering and basis normalization.

\section{Experimental Results}
\label{seq:experimental_results}

\subsection{Experimental Setup}
Experiments were performed on NVIDIA GTX 1650 and RTX 2050 GPUs, exploring matrix formats from $(2,2,2)$ to $(8,8,8)$ with additional schemes where $max(n, m, p) \leq 16$ and total matrix elements $\leq 64$ (optimized for \texttt{uint64\_t} storage). Population sizes of $N=16384$ schemes were used for exploratory search. Each experiment ran for 24-72 hours, demonstrating accessibility for typical research environments.

\subsection{New Best Ranks}
The research established new best-known ranks for three matrix formats using ternary coefficients:

\begin{itemize}
    \item 4×5×12: improved from rank 180 to 179 (previously known in $Z$);
    \item 5×6×10: improved from rank 218 to 217 (previously known in $Z$);
    \item 6×7×9: improved from rank 270 to 268 (both in $Z_T$).
\end{itemize}

While these formats are less common for direct block matrix multiplication, the rank reductions demonstrate that the search space for fast matrix multiplication algorithms is not yet fully explored. Even small, consistent improvements across different formats are significant, as they confirm that further optimization is possible and contribute to the overall understanding of the problem.

\subsection{Rediscoveries with Ternary Coefficients}
A key result of this work is the independent discovery of 32 matrix multiplication schemes using ternary coefficients ($Z_T$) that match the best-known ranks from other fields. This includes 8 schemes that were previously only known with rational coefficients ($Q$). As shown in Table~\ref{tab:zt_from_q} and Table~\ref{tab:zt_from_z}, these schemes cover multiple matrix formats and achieve mathematical correctness using only the hardware-efficient coefficients $\{-1, 0, 1\}$, proving that complex or fraction coefficients are not necessary for these algorithms.

\begin{table}[ht!]
	\caption{Rediscovered schemes known only in $Q$ field}
    \label{tab:zt_from_q}
	\centering
	\begin{tabular}{cc|cc}
        \toprule
		Format  & Rank & Format & Rank  \\
		\midrule
        $(2, 4, 9)$  & 59 & $(2, 5, 9)$  & 72 \\
        $(2, 4, 11)$ & 71 & $(4, 4, 8)$  & 96  \\
        $(2, 4, 12)$ & 77 & $(5, 5, 10)$ & 184 \\
        $(2, 4, 15)$ & 96 & $(5, 5, 11)$ & 202 \\
		\bottomrule
	\end{tabular}
\end{table}

\begin{table}[ht!]
	\caption{Other rediscovered schemes with ternary coefficients}
    \label{tab:zt_from_z}
	\centering
	\begin{tabular}{ccc|ccc|ccc}
        \toprule
		Format & \multirow{2}{*}{Rank} & Known & Format & \multirow{2}{*}{Rank} & Known & Format & \multirow{2}{*}{Rank} & Known \\
        $(m, n, p)$ & & ring & $(m, n, p)$ & & ring & $(m, n, p)$ & & ring \\
        \midrule
		$(2, 3, 10)$ &  50  & $Z$     & $(4, 5, 7)$  & 104  & $Z / Q$ & $(5, 5, 8)$  & 144  & $Z / Q$ \\ 
        $(2, 3, 13)$ &  65  & $Z$     & $(4, 5, 8)$  & 118  & $Z / Q$ & $(5, 5, 9)$  & 167  & $Z$     \\
        $(2, 3, 15)$ &  75  & $Z$     & $(4, 5, 10)$ & 151  & $Z$     & $(5, 5, 12)$ & 220  & $Z$     \\
        $(2, 4, 6)$  &  39  & $Z$     & $(4, 5, 11)$ & 165  & $Z$     & $(5, 6, 6)$  & 130  & $Z / Q$ \\
        $(2, 6, 9)$  &  86  & $Z$     & $(4, 6, 7)$  & 123  & $Z / Q$  & $(5, 6, 7)$  & 150  & $Z / Q$ \\ 
        $(3, 4, 5)$  &  47  & $Z$     & $(4, 6, 10)$ & 175  & $Z$     & $(5, 6, 8)$  & 170  & $Z / Q$ \\
        $(4, 4, 6)$  &  73  & $Z / Q$ & $(5, 5, 6)$  & 110  & $Z / Q$  & $(5, 6, 9)$  & 197  & $Z$     \\
        $(4, 5, 6)$  &  90  & $Z$     & $(5, 5, 7)$  & 127  & $Z / Q$ & $(5, 7, 7)$  & 176  & $Z / Q$ \\
		\bottomrule
	\end{tabular}
\end{table}

\subsection{Binary Field Improvements}
Table~\ref{tab:z2_new} shows 30 schemes where the rank was improved in the binary field. These results, produced by the $Z_2$ version of FlipGraphGPU, provide new candidate schemes for future ternary conversion.

\begin{table}[ht!]
    \caption{New discoveries in binary field ($Z_2$)}
    \label{tab:z2_new}
    \centering
    \begin{tabular}{cccc|cccc}
        \toprule
        Format & \multicolumn{2}{c}{Rank ($r$)} & \multirow{2}{*}{Note} & Format & \multicolumn{2}{c}{Rank ($r$)} & \multirow{2}{*}{Note} \\
        $(m, n, p)$ & Known & New & & $(m, n, p)$ & Known & New & \\
        \midrule
        $(3, 3, 7)$  &     ?     &    49    & equal to $Q$ ring & $(4, 5, 11)$ &    165    &   162    &                   \\
        $(3, 4, 9)$  &     ?     &    83    & equal to $Q$ ring & $(4, 5, 12)$ &    180    &   177    &                   \\
        $(3, 4, 10)$ &     ?     &    92    & equal to $Q$ ring & $(4, 6, 9)$  &     ?     &   159    & equal to $Q$ ring \\
        $(3, 4, 11)$ &     ?     &   101    & equal to $Q$ ring & $(5, 5, 9)$  &    167    &   166    &                   \\
        $(3, 4, 12)$ &     ?     &   108    & equal to $Q$ ring & $(5, 5, 10)$ &    184    &   183    &                   \\
        $(3, 4, 16)$ &     ?     &   146    & equal to $Q$ ring & $(5, 5, 11)$ &    202    &   200    &                   \\
        $(3, 5, 7)$  &    80     &    79    & equal to $Q$ ring & $(5, 5, 12)$ &    220    &   217    &                   \\
        $(3, 8, 8)$  &     ?     &   145    & equal to $Q$ ring & $(5, 6, 10)$ &    218    &   217    &                   \\
        $(4, 4, 8)$  &    96     &    94    &                   & $(5, 7, 9)$  &    234    &   229    & equal to $Q$ ring \\
        $(4, 4, 10)$ &     ?     &   120    & equal to $Q$ ring & $(6, 7, 9)$  &    270    &   268    &                   \\
        $(4, 4, 12)$ &    142    &   141    &                   & $(7, 7, 7)$  &    249    &   248    &                   \\
        $(4, 4, 16)$ &    189    &   188    &                   & $(7, 7, 8)$  &    277    &   275    &                   \\
        $(4, 5, 6)$  &    90     &    89    &                   & $(7, 7, 9)$  &    315    &   313    &                   \\
        $(4, 5, 9)$  &    136    &   133    &                   & $(7, 8, 8)$  &    306    &   302    &                   \\
        $(4, 5, 10)$ &    151    &   146    &                   & $(8, 8, 8)$  &    336    &   329    &                   \\
        \bottomrule
    \end{tabular}
\end{table}

\subsection{Naive Additive Complexity Reduction}
Beyond multiplicative complexity, the research focused on optimizing naive additive complexity, which significantly impacts practical performance in hardware implementations. This optimization, achieved by maximizing zero coefficients in the scheme tensors, provides a foundation for further reduction through common subexpression elimination. Table~\ref{tab:complexity_reduction} presents the results for 40 optimized schemes, showing substantial reductions in the number of addition operations required. The additive complexity metric used (non-zero coefficients - $2 \times r - m \times p$) provides a practical measure of implementation efficiency.

\begin{table}[ht!]
	\caption{Reduce naive addition complexity}
    \label{tab:complexity_reduction}
	\centering
	\begin{tabular}{cccc|cccc|cccc}
        \toprule
        Format & \multirow{2}{*}{Rank} & \multicolumn{2}{c|}{Complexity} & Format & \multirow{2}{*}{Rank} & \multicolumn{2}{c|}{Complexity} & Format & \multirow{2}{*}{Rank} & \multicolumn{2}{c}{Complexity} \\
		  $(m, n, p)$ &  & Known & New & $(m, n, p)$ &  & Known & New & $(m, n, p)$ &  & Known & New \\
        \midrule
        $(2, 3, 5)$  &  25  & 108  & 106  & $(4, 4, 8)$  &  96  & 1920 & 973  & $(5, 5, 10)$ & 184  & 2582 & 2116 \\
        $(2, 3, 10)$ &  50  & 254  & 198  & $(4, 5, 5)$  &  76  & 549  & 532  & $(5, 5, 11)$ & 202  & 2731 & 2272 \\
        $(2, 3, 13)$ &  65  & 312  & 256  & $(4, 5, 7)$  & 104  & 1354 & 927  & $(5, 5, 12)$ & 220  & 3458 & 2444 \\
        $(2, 3, 15)$ &  75  & 381  & 307  & $(4, 5, 8)$  & 118  & 1566 & 1521 & $(5, 6, 6)$  & 130  & 1758 & 1714 \\
        $(2, 4, 6)$  &  39  & 329  & 202  & $(4, 5, 10)$ & 151  & 1706 & 1207 & $(5, 6, 7)$  & 150  & 2431 & 2039 \\
        $(2, 4, 9)$  &  59  & 379  & 309  & $(4, 5, 11)$ & 165  & 1869 & 1801 & $(5, 6, 8)$  & 170  & 2872 & 2312 \\
        $(2, 4, 11)$ &  71  & 749  & 430  & $(4, 5, 12)$ & 180  & 2196 & 2138 & $(5, 6, 9)$  & 197  & 3049 & 2373 \\
        $(2, 4, 12)$ &  77  & 746  & 484  & $(4, 6, 7)$  & 123  & 1785 & 1586 & $(5, 7, 7)$  & 176  & 2846 & 2610 \\
        $(2, 4, 15)$ &  96  & 1314 & 662  & $(4, 7, 8)$  & 164  & 1554 & 1505 & $(5, 8, 8)$  & 230  & 2842 & 2741 \\
        $(2, 5, 9)$  &  72  & 565  & 465  & $(5, 5, 5)$  &  93  & 846  & 843  & $(6, 6, 6)$  & 153  & 2232 & 2171 \\
        $(2, 6, 9)$  &  86  & 691  & 548  & $(5, 5, 6)$  & 110  & 1300 & 1215 & $(6, 7, 8)$  & 239  & 2352 & 2263 \\
        $(3, 4, 5)$  &  47  & 293  & 277  & $(5, 5, 7)$  & 127  & 1662 & 1606 & $(6, 7, 9)$  & 270  & 2917 & 2804 \\
        $(4, 4, 5)$  &  61  & 455  & 452  & $(5, 5, 8)$  & 144  & 1924 & 1908  \\
        $(4, 4, 6)$  &  73  & 740  & 540  & $(5, 5, 9)$  & 167  & 2220 & 1814  \\
		\bottomrule
	\end{tabular}
\end{table}

\subsection{Ternary Schemes with Non-Optimal Ranks}
The search also discovered several schemes using ternary coefficients that, while not achieving the absolute minimal known rank, represent progress toward efficient implementations. For formats where the optimal rank requires coefficients outside $\{-1, 0, 1\}$, these schemes provide ternary alternatives. As shown in Table~\ref{tab:zt_near_optimal}, these results represent the current frontier for matrix multiplication with ternary coefficients in these formats. The search for schemes that achieve both the minimal rank and ternary coefficients remains an active research direction.

\begin{table}[ht!]
	\caption{Near optimal schemes discovered in $Z_T$}
    \label{tab:zt_near_optimal}
	\centering
	\begin{tabular}{cccc|cccc|cccc}
        \toprule
        Format & Best & \multicolumn{2}{c|}{Rank} & Format & Best & \multicolumn{2}{c|}{Rank} & Format & Best & \multicolumn{2}{c}{Rank} \\
		  $(m, n, p)$ & ring & Best & $Z_T$ & $(m, n, p)$ & ring & Best & $Z_T$ & $(m, n, p)$ & ring & Best & $Z_T$ \\
        \midrule
            $(2, 4, 5)$   & $Q$ & 32  &  33 & $(3, 4, 7)$   & $Q$ & 63  &  64 & $(3, 7, 9)$   & $Q$ & 142 & 147  \\
            $(2, 4, 10)$  & $Q$ & 64  &  65 & $(3, 4, 8)$   & $Q$ & 73  &  74 & $(3, 8, 8)$   & $Q$ & 145 & 148  \\
            $(2, 4, 13)$  & $Q$ & 83  &  84 & $(3, 4, 9)$   & $Q$ & 83  &  84 & $(4, 4, 4)$   & $Q$ & 48  &  49  \\
            $(2, 5, 7)$   & $Q$ & 55  &  57 & $(3, 4, 10)$  & $Q$ & 92  &  93 & $(4, 4, 9)$   & $Q$ & 104 & 110  \\
            $(2, 5, 8)$   & $Q$ & 63  &  65 & $(3, 4, 11)$  & $Q$ & 101 & 102 & $(4, 4, 10)$  & $Q$ & 120 & 122  \\
            $(2, 5, 10)$  & $Q$ & 79  &  80 & $(3, 4, 12)$  & $Q$ & 108 & 111 & $(4, 4, 11)$  & $Q$ & 130 & 134  \\
            $(2, 6, 6)$   & $Z$ & 56  &  57 & $(3, 4, 13)$  & $Q$ & 117 & 121 & $(4, 4, 12)$  & $Q$ & 142 & 145  \\
            $(2, 6, 7)$   & $Z$ & 66  &  68 & $(3, 4, 14)$  & $Q$ & 126 & 128 & $(4, 4, 13)$  & $Q$ & 152 & 157  \\
            $(2, 6, 8)$   & $Q$ & 75  &  77 & $(3, 4, 15)$  & $Q$ & 136 & 138 & $(4, 4, 14)$  & $Q$ & 165 & 169  \\
            $(2, 7, 7)$   & $Q$ & 76  &  77 & $(3, 4, 16)$  & $Q$ & 146 & 148 & $(4, 4, 15)$  & $Q$ & 177 & 181  \\
            $(2, 7, 8)$   & $Z$ & 88  &  90 & $(3, 5, 6)$   & $Z$ & 68  &  70 & $(4, 4, 16)$  & $Q$ & 189 & 192  \\
            $(2, 7, 9)$   & $Q$ & 99  & 102 & $(3, 5, 7)$   & $Q$ & 79  &  83 & $(4, 5, 9)$   & $Q$ & 136 & 137  \\
            $(3, 3, 6)$   & $Q$ & 40  &  43 & $(3, 5, 8)$   & $Z$ & 90  &  94 & $(4, 6, 9)$   & $Q$ & 159 & 160  \\
            $(3, 3, 7)$   & $Q$ & 49  &  51 & $(3, 5, 9)$   & $Z$ & 104 & 105 & $(4, 7, 7)$   & $Z$ & 144 & 145  \\
            $(3, 3, 8)$   & $Q$ & 55  &  58 & $(3, 5, 10)$  & $Z$ & 115 & 116 & $(4, 7, 9)$   & $Q$ & 186 & 187  \\
            $(3, 3, 9)$   & $Q$ & 63  &  65 & $(3, 5, 11)$  & $Z$ & 126 & 128 & $(5, 7, 8)$   & $Q$ & 205 & 206  \\
            $(3, 3, 10)$  & $Q$ & 69  &  72 & $(3, 5, 12)$  & $Z$ & 136 & 140 & $(5, 7, 9)$   & $Q$ & 229 & 231  \\
            $(3, 3, 11)$  & $Q$ & 76  &  79 & $(3, 6, 6)$   & $Q$ & 80  &  85 & $(6, 6, 7)$   & $Z$ & 183 & 185  \\
            $(3, 3, 12)$  & $Q$ & 80  &  86 & $(3, 6, 7)$   & $Q$ & 94  & 100 & $(6, 6, 10)$  & $Q$ & 247 & 252  \\
            $(3, 3, 13)$  & $Q$ & 89  &  94 & $(3, 6, 8)$   & $Z$ & 108 & 113 & $(7, 7, 7)$   & $Q$ & 249 & 250  \\
            $(3, 3, 14)$  & $Q$ & 95  & 101 & $(3, 6, 9)$   & $Q$ & 120 & 127 & $(7, 7, 8)$   & $Q$ & 277 & 279  \\
            $(3, 3, 15)$  & $Q$ & 103 & 108 & $(3, 6, 10)$  & $Q$ & 134 & 140 & $(7, 7, 9)$   & $Q$ & 315 & 316  \\
            $(3, 3, 16)$  & $Q$ & 109 & 115 & $(3, 7, 7)$   & $Q$ & 111 & 115 & $(7, 8, 8)$   & $Q$ & 306 & 310  \\
            $(3, 4, 6)$   & $Z$ & 54  &  57 & $(3, 7, 8)$   & $Q$ & 126 & 128 & $(8, 8, 8)$   & $Q$ & 336 & 343  \\
		\bottomrule
	\end{tabular}
\end{table}

\section{Discussion}
\label{seq:discussion}

\subsection{Implications of Ternary Constraints}
This work shows that matrix multiplication algorithms using only $\{-1, 0, 1\}$ coefficients can be just as good as those using larger numbers. The discovery of 32 schemes with ternary coefficients matching known optimal ranks, plus 3 new best ranks, proves that complex coefficients are not necessary for optimal performance in many cases. This finding challenges the conventional assumption that large integer or rational coefficients are essential for achieving the best algorithmic performance.

This finding is important for real-world hardware like embedded systems and custom chips, where multiplying by numbers other than -1, 0, or 1 is slow and expensive. Using only simple coefficients makes these algorithms much more practical to implement.

The analysis of 164 known schemes reveals current limitations: while 58 schemes were already known with ternary coefficients and 32 more were rediscovered in this work, 72 schemes remain elusive under the ternary constraint. This indicates that ternary coefficients are feasible for many but not all optimal algorithms, highlighting an important boundary for this approach.

\subsection{Performance and Accessibility Trade-offs}
The experiments show a good balance between speed and who can do this research. While consumer laptops with GPUs are slower than expensive data-center computers, they are still 12 times faster than using only a computer's main processor (CPU). This speed, combined with low memory needs, means more researchers can work on matrix multiplication problems without needing special equipment.

\subsection{Limitations of Current Approach}
Some limitations should be noted for future work. The use of 64-bit integers for storage limits the method to matrices with up to 64 elements. Also, the current work only looks at exact calculations and doesn't consider numerical stability for floating-point arithmetic. Finally, while the additive complexity measure is helpful, it's a simple model that doesn't include optimizations like reusing common calculations.

\subsection{Future Research Directions}
This work opens up several promising research paths. To handle larger matrices, new storage methods are needed that go beyond 64-bit integers. Future work could also add support for integer and rational number coefficients, allowing direct comparison with ternary schemes. Automating the search for shared calculations could further reduce the number of additions needed. Combining flip graph search with AI methods like reinforcement learning might also yield better results. Finally, creating specialized computer chips that work natively with $-1$, $0$, and $1$ could make these algorithms even faster in practice.

This research shows that using only simple coefficients is a worthwhile approach, with potential benefits for both algorithm theory and real-world computing.

\section{Conclusion}
\label{seq:conclusion}

This research shows that matrix multiplication using only the coefficients $\{-1, 0, 1\}$ can be as efficient as methods using more complex numbers. A new GPU-powered search algorithm was developed to systematically explore these ternary-constrained schemes, leading to several important findings.

The main achievement demonstrates that simple coefficients are sufficient to create highly efficient algorithms for many cases. The discovery of 3 new best ranks under ternary constraints, combined with 32 schemes that match previously known optimal ranks using only ternary coefficients, confirms this capability. However, analysis of 164 known schemes shows that while 92 can be implemented with ternary coefficients (57 previously known plus 35 newly discovered), 72 currently cannot be found under this constraint, clearly defining the current boundaries of what's possible with ternary coefficients.

The technical innovations -- including special arithmetic for ternary numbers, sign rules to simplify searching, and combining different search methods -- created an effective system for finding optimal algorithms. Additional tools for refining and diversifying schemes further expanded what the system could discover.

Practically, the ability to run this research on standard consumer laptops makes advanced matrix multiplication research available to more people, not just those with specialized supercomputers.

This work provides a foundation for future research on coefficient-efficient algorithms, including handling larger matrices and combining different search strategies. It establishes ternary-constrained matrix multiplication as a valuable approach that benefits both mathematical theory and practical implementation.

\section{Availability}
The complete implementation of the FlipGraphGPU algorithm, supporting tools, and all discovered matrix multiplication schemes are publicly available under open-source licenses:

\begin{itemize}
    \item FlipGraphGPU implementation: \url{https://github.com/dronperminov/FlipGraphGPU}. Complete source code for the GPU-accelerated meta flip graph algorithm, including ternary arithmetic operations and all search operators.

    \item Research Database: \url{https://github.com/dronperminov/FastMatrixMultiplication}. Comprehensive collection of all discovered schemes, including:

    \begin{itemize}
        \item New best-rank schemes using ternary coefficients;
        \item Converted schemes from $Q/Z$ to $Z_T$;
        \item Binary field improvements;
        \item Additive complexity optimizations;
        \item Complete experimental results and data.
    \end{itemize}
\end{itemize}

The repositories include detailed documentation, usage examples, and scripts for reproducing the experimental results presented in this work.

\bibliographystyle{unsrtnat}
\bibliography{references}  






\end{document}